# Self-sweeping ytterbium-doped fiber laser based on a fiber saturable absorber

Zengrun Wen[1,2,3], Kaile Wang[1,2,3], Baole Lu[1,2*], Haowei Chen[1,2], and Jintao Bai[1,2*]

[1]*National Key Laboratory of Western Energy Photonics Technology, International Joint Research Center on Photoelectric Technology and Functional Nanomaterials, Institute of Photonics & Photon-technology, Northwest University, Xi′an 710069, China*
[2]*Shaanxi Engineering Technology Research Center for Solid State Lasers and Application, Provincial Key Laboratory of Photo-electronic Technology, Northwest University, Xi′an 710069, China*
[3]*These authors contributed equally to this work*
*E-mail: lubaole1123@163.com, baijt@nwu.edu.cn

**Abstract:** Generally speaking, the self-sweeping effect relies on the dynamical grating formed in a gain fiber. Here, the normal self-sweeping was generated in a pump-free ytterbium-doped fiber which serves as a fiber saturable absorber and is introduced to the laser cavity by a circulator in this experiment. The sweeping rate and the sweeping range alter as usual, both of which can be controlled by the pump power. Further, a new self-pulse signal is observed and discussed in this work, which shows the difference of the self-sweeping effects between active fiber and fiber saturable absorber.

Tunable laser sources are widely used in a variety of applications such as optical communication, coherent beam combining and optical sensors [1-3]. The proverbial methods that generate tunable light sources rely on the grating and band-pass filter. One of the special tunable sources is wavelength-sweeping fiber laser, in which the central wavelength can be tuned periodically. To achieve the periodical and stable tuning operation, the active sweeping devices such as electric-driven PZT, heater or the scanning tunable filter have been utilized in the majority of modern lasers [4-8]. In recent years, a kind of spontaneous sweeping fiber laser based on so-called self-sweeping or the self-induced laser line sweeping effect was found and deeply studied. The self-sweeping regime that a spontaneous, periodical, stable tuning running can be obtained by the spatial burning hole of the active medium in a stand-wave condition. In 2011, self-sweeping fiber laser (SSFL) was reported in ytterbium-doped fiber (YDF) laser based on the non-Fabry-Perot cavity, which revealed a new wavelength-tunable phenomenon from shorter to longer wavelengths without tunable elements. Describing in detail the generation and annihilation in the self-sweeping regime later [9,10]. SSFL reported only in YDF pushes the researcher to attempt in different doped fibers [11-13]. Subsequently, neodymium-, bismuth-, erbium-, thulium-holmium-, holmium-, and thulium-doped fiber serve as active media, all of which can obtain the self-sweeping regime with broadly sweeping ranges [14-20]. Besides, the self-sweeping fiber laser with single frequency signals displays the remarkable features in the spectrum and intensity dynamics. Therefore, SSFLs have actively promoted the development of laser spectroscope, spectrum of water absorption lines, the generation of ultra-short pulse and optical fiber sensor. Nowadays, the research on the self-sweeping effect has involved many phenomena. The sweeping direction can be defined by the normal self-sweeping and reverse self-sweeping in which the generated wavelength increases and decreases in the sweeping range, respectively [21]. The self-sweeping fiber laser in the linear cavity has produced multiple sweeping directions involving normal self-sweeping, reverse self-sweeping and a mixed state [22,23]. The up-to-date bi-directional fiber ring lasers have reported for the generation of reverse self-sweeping effect in Tm and Yb active fiber [19,24,25].

The reported fiber lasers whatever in linear cavities or the bidirectional fiber ring cavities, the obvious dependence on self-sweeping effect on the standing-wave condition to allow us to make an attempt for forming a self-induced grating in fiber saturable absorber (FSA) [26,27]. In this Letter, we report a self-sweeping effect by generating a dynamics grating in pump-free ytterbium-doped fiber (PFYDF). A new self-pulse phenomenon is observed in temporal dynamic measurement, which reveals the building and duration of dynamics induced grating. Besides, the fiber laser generates a normal sweeping with the largest coverage of 4.85 nm. These results will extend the knowledge and provide a new implementation method for the self-sweeping fiber laser.

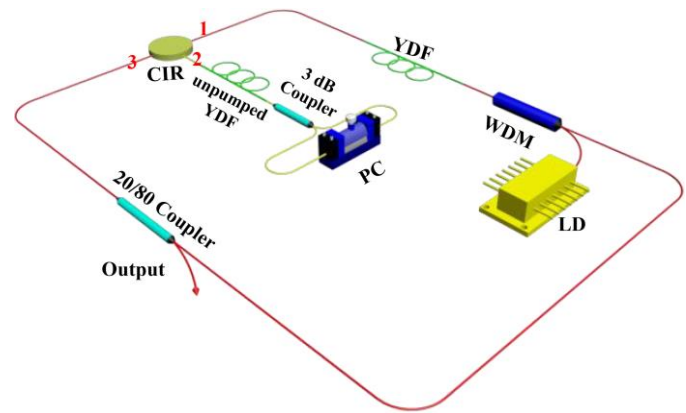

Fig. 1. Experimental setup of the proposed wavelength self-sweeping Yb-doped fiber laser.

The schematic diagram of the proposed SSFL is illustrated in Fig. 1. The active medium is pumped by a 975 nm laser diode (LD) connected with a 975/1060 nm wavelength division multiplexer (WDM). In the present scheme, a 1 m-long single-mode Yb-doped fiber (CorActive Yb 501) is used as an active medium. A circulator (CIR) is employed to introduce the 1.8 m YDF and fiber loop mirror (FLM). Besides, the CIR can ensure the unidirectional operation in the laser because of the high reverse isolation. The 20/80 coupler is severed as the output coupler for measurement. The FLM consists of a 3 dB coupler whose two ports of output are spliced fusion. Note that this simple cavity can not offer any narrowband wavelength elements. Therefore, the appearance of central wavelength only depends on the dynamics induced grating in the PFYDF. The polarization controller (PC) in the cavity assures a better result in the experiment.

In our experiment, the phenomenon of self-sweeping can be observed over the threshold pump power of 65 mW. With the increase of pump power, the self-sweeping of wavelength emerges obviously, and self-pulse is observed simultaneously. When the pump power is up to 260 mW, explosive chaotic spectrum borns and self-pulse disappears immediately. In measurement, the output spectra are observed by the optical spectral analyzer (OSA Yokogawa, AQ6370C). Figure 2 shows the optical spectra information of the self-sweeping phenomena. In Figs. 2(a) and (b), the periodical normal sweeping of wavelength in the range of 4.83 nm with a rate of 0.24 nm/s at the pump power of 180 mW is depicted in detail. Besides, there is a relationship of square root function between sweeping rate α [nm/s] and output power P [mW] which is obtained through fitting [see Fig. 2(c)]. Figure 2(d) depicts that the sweeping range changes with the increasing output power. All the above observed spectral behaviors are similar to those in the self-sweeping effect of active fiber.

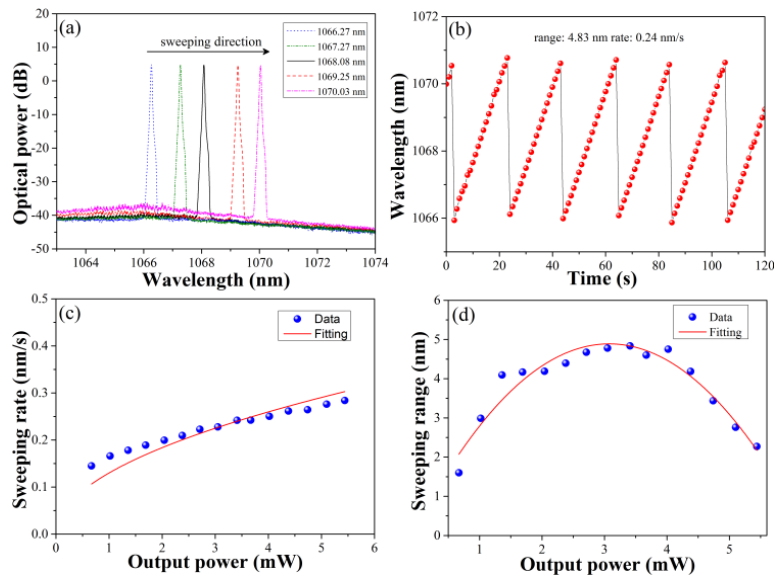

Fig. 2. (a) and (b) Wavelength dynamics measured by an OSA at a pump power of 180 mW. (c) Average sweeping rate as a function of output power. (d) Sweep range as a function of output power.

The temporal dynamics of the proposed fiber laser can be monitored by a detector (DET08CFC Thorlabs) and an oscilloscope (Tektronix DPO7254C). The intensity signal is displayed in Fig. 3. Figures 3(a) and (b) exhibit the pulse signal of stable self-sweeping at 100 mW and 220 mW, respectively. While the pump power is large enough, the unstable and chaotic temporal dynamics emerge, which are shown in Figs. 3(c) and (d). Consistent with all self-sweeping lasers, the stable self-sweeping running will be terminated with the emergence of disordered pulse signals [Fig. 3(c) and (d)]. In previous articles, the dynamics phase and gain gratings formed in an active medium.

Owing to the laser running in the regimes of few longitudinal modes, the pulses are modulated with the inter-mode beating frequency and associated with self-sustained relaxation oscillatio. Such a self-pulse regime exhibits regular microsecond pulses and irregular peak pulses. Differ with the microsecond self-pulsing signal, we observed a new self-pulse signal, as shown in Fig. 3(a), the pulse dynamical signal under the stable self-sweeping running exhibits periodicity and the continuous wave generates between the pulses. We contribute that the switch between pulse and continuous states represents the shifting process of the central wavelength.

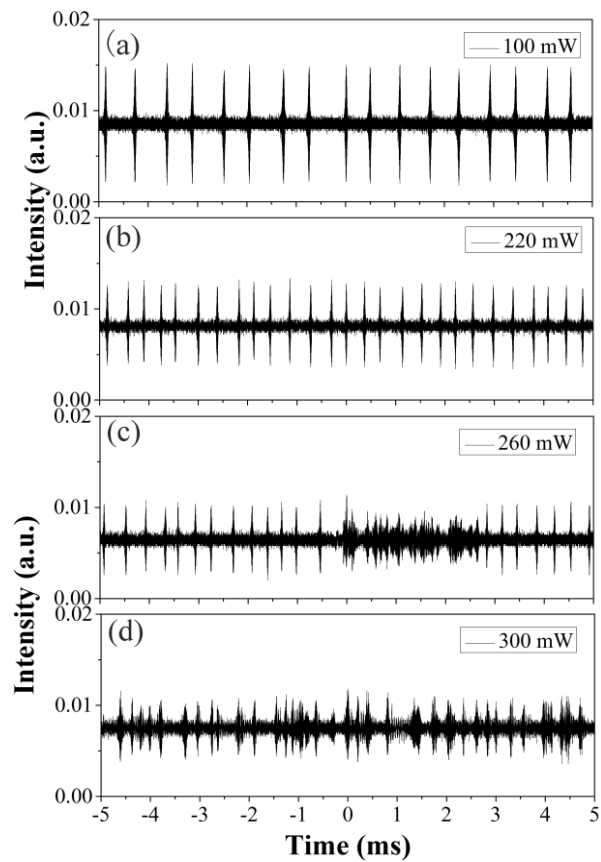

Fig. 3. The intensity dynamics of fiber laser. (a) Stable self-sweeping operation at a pump power of 100 mW. (b) Stable self-sweeping operation at a pump power of 220 mW. (c) Unstable self-sweeping operation at a pump power of 260 mW. (d) Chaotic spectrum running at a pump power of 300 mW.

Figure 4 shows the variation of the average pulse repetition rate which increases with the increasing output power. When the pump power is increased, the average repetition rate increases from 1.5 kHz to 2.7 kHz with a tendency of square root function versus the output power. One can see that the trend of pulse repetition rate is coincident with the sweeping rate, which manifests the change of central wavelength is synchronized with the reproduction of the pulse. Besides, the behavior can be related to the lifetime of dynamic gratings according to the published work, in which the relaxation rate of recording grating in the Yb-doped fiber reduces with an ascending input power [28]. Therefore, the behavior of pulse also embodies the grating features.

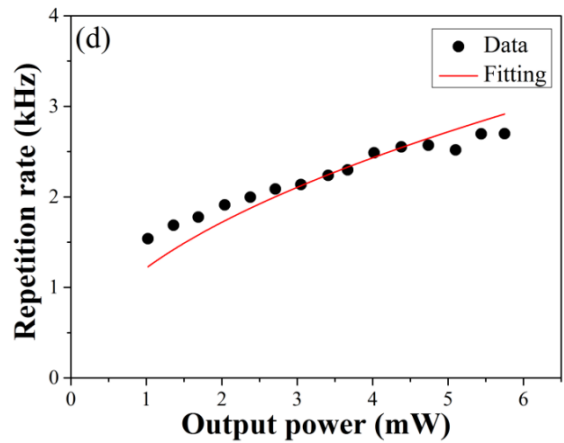

Fig. 4. Average repetition rate as a function of output power.

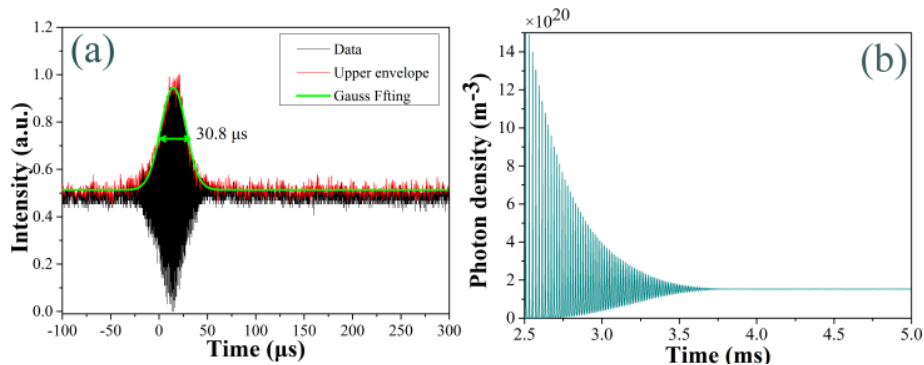

Fig. 5. (a). Intensity dynamics of single interference signal at the pump power of 100 mW. (b) the CW operation at low population density and long decay time of SA.

Figure 5(a) reveals the observed single temporal pulse of this fiber laser in a small time window. The duration time is 30.8 μs at the pump power of 100 mW. We can see that the laser output before and after the duration of the pulse signal is the continuous wave (CW), namely, the laser frequency is changed during the signal duration. Further, the duration of the pulse signal can be considered as the building time of dynamics induce gratings in the YDFSA. The appearance of the pulse signal changes the output wavelength of this laser. The further pulse analysis based on rate equation can be considered to the feature of SA. In general, the stable and unstable regimes can be obtained in different conditions [29]. The population density of the upper level of absorber states plays an important role. When a section of YDF acts as the saturable absorber in a fiber laser, the low population density and response time of FSA lead to the CW operation. The photon density of our fiber laser depicted in Fig. 5(b) which explains the damped pulse dynamics. Therefore, the entire process of intensity dynamics shows both the feature of SA and grating. As for the self-sweeping in active fiber, the population density of the upper level was excited by signal light and pump light. The whole population of the upper level devotes to the light amplification which expands the population of absorber state and further cut down the lifetime of the dynamic grating.

In summary, the effect of self-sweeping was observed in a fiber laser with a YDFSA. The damped pulse can be measured when the fiber laser operating in the self-sweeping regime, which reveals a different drive model of wavelength self-sweeping effect. The change of laser frequency depends on the dynamics induced grating produced by the light interference in YDFSA. The repetition rate shows the change of laser frequency, which is consistent with the sweeping rate, both of which increase with output power as a square root function. In general, the behaviors of our laser such as the process of coming to an end, the pattern of sweeping rate and range is the same as the previously reported self-sweeping fiber laser, the only difference is temporal dynamics. Generally, the pump-free doped fibers are used to achieve single-frequency running. Therefore, our work expands the application of pump-free doped fibers and promote the control of lasing in fiber lasers.

## Acknowledgments

This work is supported by National Nature Science Foundation of China (61905193); National Key R&D Program of China (2017YFB0405102); Key Laboratory of Photoelectron of Education Committee Shaanxi Province of China (18JS113); Key R&D project of Shaanxi Province-International Science and Technology Cooperation Project (2020KW-018).